**Title:** Using mathematics to study how people influence each other's opinions

**Authors:** Grace J. Li, Jiajie Luo, Kaiyan Peng, and Mason A. Porter

## Abstract

People sometimes change their opinions when they discuss things with other people. Researchers can use mathematics to study opinion changes in simplifications of real-life situations. These simplified settings, which are examples of mathematical models, help researchers explore how people influence each other through their social interactions. In today's digital world, these models can help us learn how to promote the spread of accurate information and reduce the spread of inaccurate information. In this article, we discuss a simple mathematical model of opinion changes that arise from social interactions. We briefly describe what such opinion models can tell us and how researchers try to make them more realistic.

## How do researchers study opinion changes?

In the short novel *Ad Eternum* by Elizabeth Bear, a character comments, "Opinions are like kittens. People are always giving them away." Everybody has opinions. Our opinions about things change over time, and they are often influenced by other people's opinions. Many researchers study how opinions change with time and how people influence each other [1,2]. They often use mathematics to investigate these ideas and gain insights into how people change their views, how groups of people form shared opinions, and how these shared opinions shift.

Imagine that your school is redesigning its logo and colors. The school principal has announced that the administration has selected two possible colors—red and blue—and that students can vote on which of those two colors will become the school's main color. During the lunch recess, you and your friends are sitting at a table and discussing the new color options. Suppose that only two people discuss the color options with each other at a time. Also suppose that your preference is red. You turn to one of your friends to discuss color preferences. Your friend says that blue would look better on school shirts and convinces you to prefer blue. Afterwards, another friend turns to you to discuss things, and you convince this friend to also prefer blue. Imagine that these pairwise discussions continue until eventually everybody at the lunch table is convinced to prefer the same color.

We can study the above discussion process using a **mathematical model** [3]. Researchers use mathematical models to investigate how opinions change due to social interactions, such as those in the students' lunch conversations. Researchers often study simple models that are easy to understand and explore. One model that researchers use for situations like the one above is called a **voter model** (which, despite its name, has almost nothing to do with voting) [3]. Voter models have two ingredients: (1) people's initial opinions, which is what they think before discussing their views with others; and (2) an **update rule**, which describes how people's opinions change. In our example, the initial opinions are the colors (red or blue) that each person initially prefers. For the update rule in this voter model, we randomly choose two people and then randomly choose one of the opinions for the two people to agree on after their discussion. Each of the two opinions is equally likely. (Imagine flipping a coin to determine which opinion to agree on.) If both people already have the same opinion, in this simple model, we can still randomly choose them to discuss their opinions with each other. However, in that case, their opinions both stay the same.  This update rule mimics a discussion in which one person can convince another person to change their color preference. This simple update rule does not capture the many details of how people interact in real life. (For example, this rule includes no memory of any previous discussions.) Nevertheless, despite the rule's naive nature, it still helps researchers study how people's opinions change with time.

In **Figure 1**, you can see how a voter model relates to our example of friends discussing their preferred school colors at lunch. When everybody eventually agrees on the same color, they have reached a **consensus**. When people do not reach a consensus, we say that they have **fragmented opinions**. For example, when people are polarized, with two different opinions (such as red or blue for the school color), their opinions are fragmented. For a voter model, researchers are interested in whether people reach consensus and, if so, how long it takes to reach it [2].

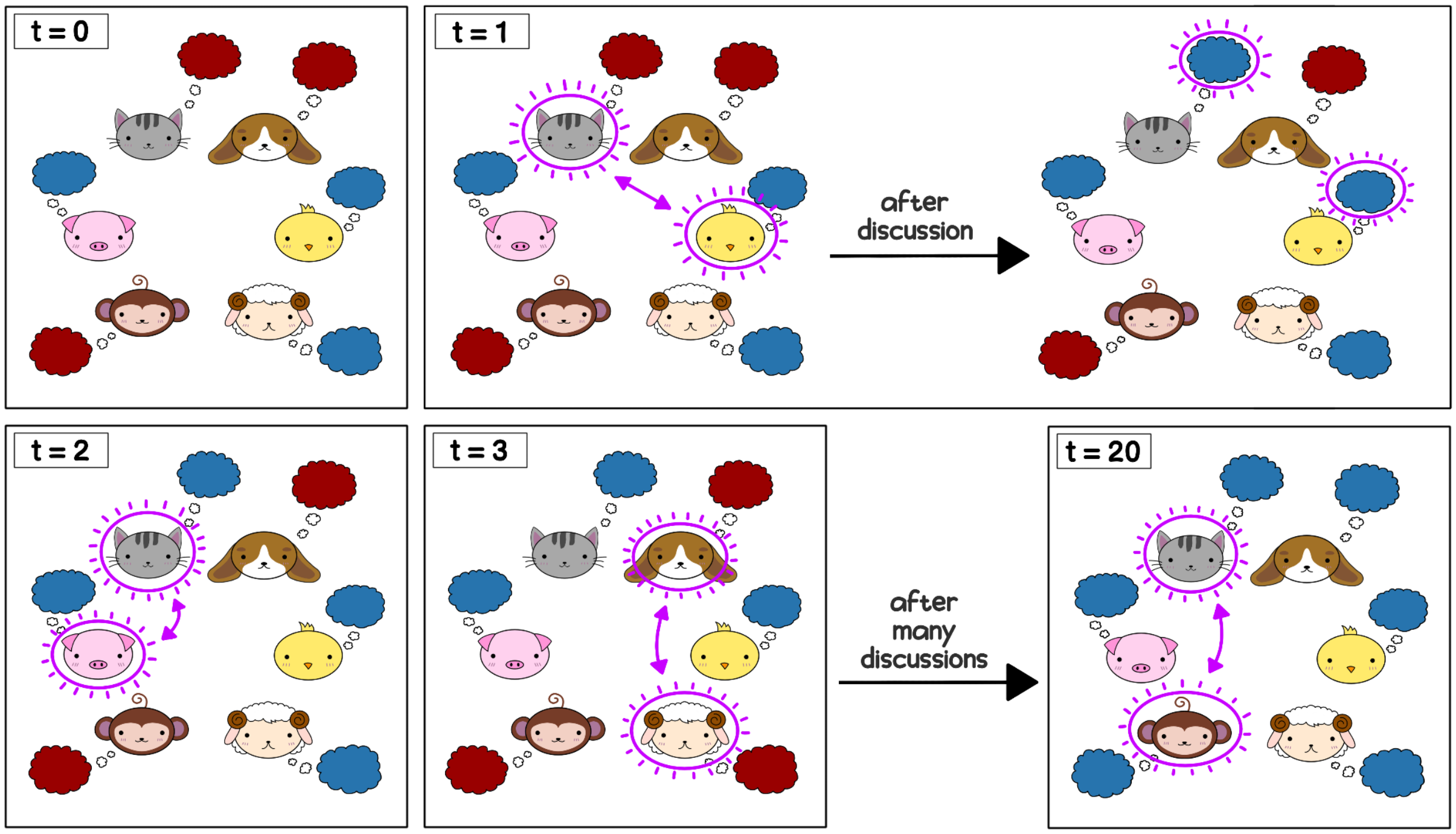

**Figure 1.** Modeling color choices using a voter model. Initially (at time t = 0), every animal prefers either blue or red for the main school color. At time t = 1, Bird and Cat discuss their preferences and decide that they prefer blue. At time t = 2, Cat and Pig discuss their preferences. They have the same opinion, so their opinions do not change. At time t = 3, Dog and Sheep discuss the school colors and decide that they prefer red. The opinion updates

continue until the animals reach a consensus. In this example, at time t = 20, everyone prefers blue.

### Incorporating social networks into opinion models

In our school-color example, a small group of friends discuss their color preferences in pairs. Suppose that these discussions occur in a large school where the students do not all know one another. Researchers can represent the collection of social relationships in a school as a **social network** of students. The students are the **nodes** of the social network, and the social relationships that connect the students are the **edges** of the network. If two people have a social relationship, we say that they are "neighbors" in the network. In our example, nodes have edges between them if they are friends. When researchers study a voter model on a social network, only neighboring nodes can influence each other's opinions. That is, they assume that only people with a social relationship influence each other. For example, students in a school tend to influence their friends. Using social networks lets researchers study how network structure—that is, who is friends with whom—influences opinion changes.

Consider a social network of friendships between students in a school. Let's look at our voter model on this social network. As before, each student initially prefers either red or blue. However, unlike in our example without a social network, some people are not friends with each other. Two people only discuss their preferences with each other if they are friends. At each update of the model, we randomly choose a pair of friends to talk to each other; that is, we choose neighbors in the social network of students. We then randomly choose one of their opinions and suppose that both students agree on this opinion after their discussion. As before, this opinion update represents one person convincing their friend to prefer the same color as them. In **Figure 2**, you can see an example of how this voter model updates on a very small social network. If you look at how the opinions change with time, eventually every student agrees to vote for the same color. That is, the students have reached a consensus. Unfortunately, in real life, it may take a very long time to reach a consensus; a school may not wait that long. To examine situations like this, researchers study how the opinions in a voter model change in a specified amount of time (such as a week) to see if a consensus emerges. Researchers also study how long it takes for a consensus to emerge.

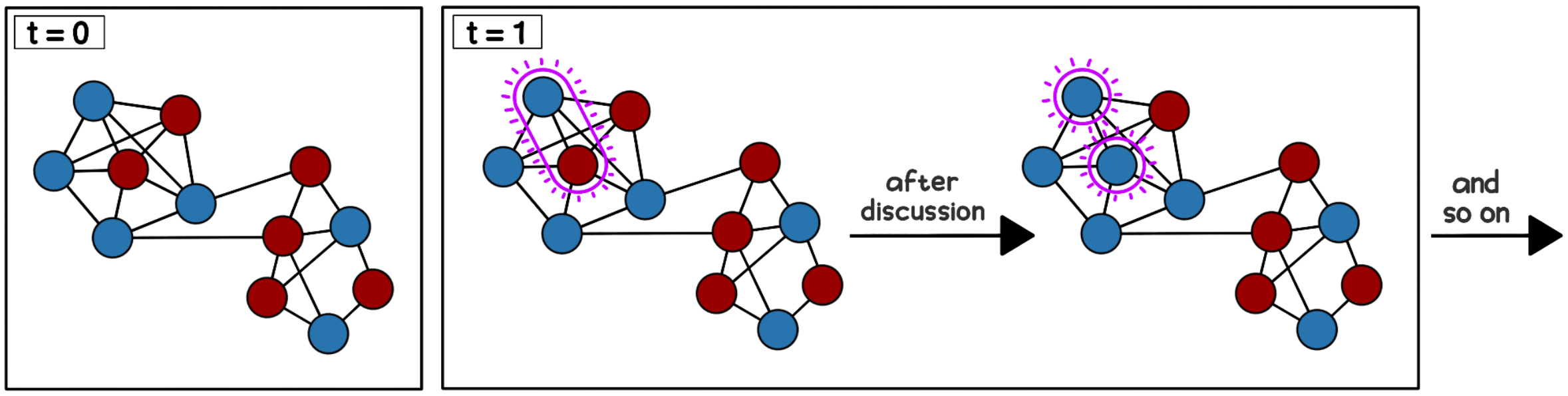

**Figure 2.** An example of how opinions can update in a voter model on a social network. The people (the nodes) are filled circles and their friendships (the edges) are black lines that connect

the nodes. We color the nodes by their opinion (blue or red). At time t = 0, we show the social network and everybody's initial opinions. Unlike in Figure 1, only friends (that is, neighboring nodes) can interact; some people are friends with each other, but others are not. At each time, we randomly select two friends to interact and update their opinions.

When studying a voter model, one possible choice is to randomly determine the initial node opinions. However, in our example, the initial color preferences probably do not come from a random process like flipping a coin. Many people prefer certain colors and dislike other colors. These opinions can arise for many reasons. Perhaps somebody often wears blue clothes or has a favorite sports team with blue as a team color. A person may also want to avoid a color that matches the color of a rival sports team. In a school social network, there are communities of people with many friendships within the community. For example, friendship communities can arise from classes, school clubs, or sports teams. People in the same community are more likely to be friends with each other than people in different communities. As you can see in **Figure 3**, researchers can choose initial opinions based on such communities. For example, suppose that the team color of the local football team is blue. Maybe more 7th-grade students than 8th-grade students are fans of the team, so perhaps 7th graders are more likely than 8th graders to prefer blue as the main school color. Researchers study how communities in a social network affect the outcome of voter models. For example, how do communities influence the opinions of people in a social network? Will the people in a social network eventually achieve consensus? If so, how long will it take?

Opinion models often include randomness. In a voter model, randomness comes from choosing the initial opinions and determining which specific pair of people interact at each time. In **Figure 3**, you can see that randomness can cause the exact same mathematical model to give different outcomes. Researchers need to be careful when figuring out whether differences in model outcomes arise from features (such as communities) of social networks or from randomness. They often use computer simulations to study opinion models, and it is important to simulate a model (such as a voter model) many times to carefully understand its possible outcomes.

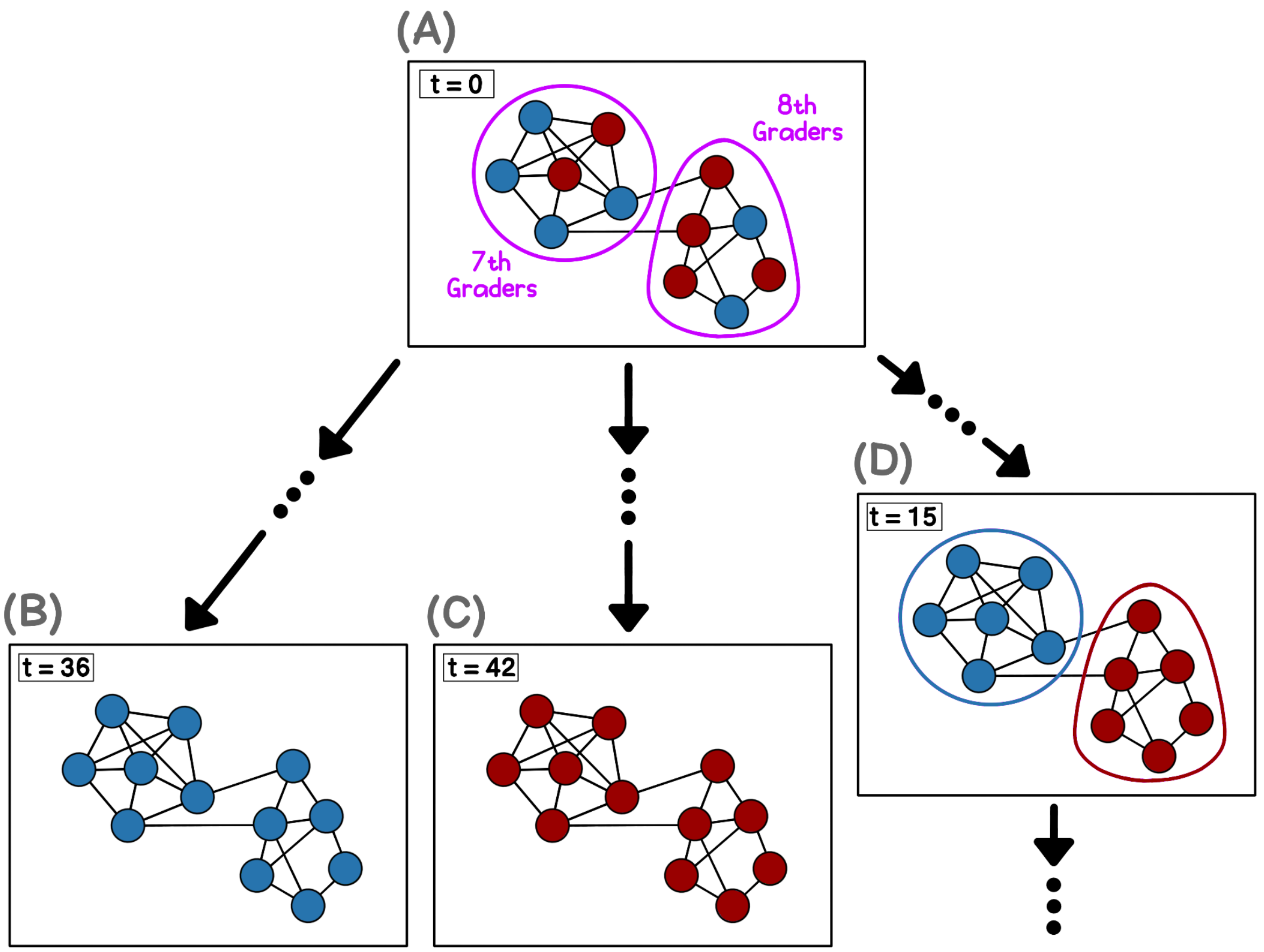

**Figure 3.** Randomly determining which nodes interact can lead to different outcomes. Consider a social network with two communities—7th graders and 8th graders—that tend to prefer blue and red, respectively. **(A)** At time t = 0, we select which color each student prefers. Using these same initial preferences, we simulate the voter model many times on a computer. **(B)** In one simulation, the students eventually all prefer blue. **(C)** In another simulation, the students eventually all prefer red. **(D)** In another simulation, the students still have different opinions at time t = 15. We continue this simulation until they reach a consensus.

**What can we learn from voter models?**

Why are researchers interested in voter models and other opinion models? Deciding on new school colors may not seem like a big deal. However, studying opinion models can also help researchers examine changes in public opinions on important topics like presidential elections and vaccination policies. Opinion models can give insights into how to promote the spread of accurate information and how the use of social media can affect our views [4]. Social media has changed how people interact and how opinions spread. For example, it has influenced people's views about COVID-19 [4]. Social media can also help health officials share important guidance, such as information about vaccines. Unfortunately, social media can also help spread false and misleading information, which can harm the mental and physical health of many people.

Researchers in many subjects—including mathematics, psychology, sociology, biology, physics, economics, computer science, and others—study opinion models for various reasons [5]. Mathematicians and physicists often study these models because they are interesting. Political scientists have used opinion models to study polarization and voting outcomes in elections [1,5]. In business, opinion models have been used to study decision-making, price-setting in financial markets, product reviews in online sales, and the effects of advertising campaigns [6]. Researchers have also examined how opinion changes and disease spread affect each other by studying models that combine them [4].

Voter models have many applications, and we can make them more useful by making them more realistic. How can we make voter models more realistic? In a voter model, we randomly select which opinion is shared by two individuals after they interact. However, in reality, it is not clear precisely how people form and change their opinions. People's opinions are also influenced by factors other than direct influence from other people. Forming and changing opinions are complicated processes, so they are hard to model mathematically. The update rule of our voter model oversimplifies reality, so researchers have tried to make such rules more realistic [1,3]. Researchers have also developed many other types of opinion models—such as models that include stubborn people who are unlikely to change their opinions—to incorporate various ideas of how people change their opinions [1,2]. They have also developed opinion models that incorporate notions of peer pressure [3]. For example, perhaps some people only change their opinion to one that is sufficiently popular among their friends (for example, when at least five of their friends have that opinion). Researchers also study opinion models in which social-network structure and opinions both change in ways that depend on each other [7]. For example, perhaps you stop following a person on TikTok or Instagram if you disagree strongly enough on an important issue.

**Summary and where to go from here**

Studying opinions and how they change is a fascinating way to apply mathematics to study how social interactions between people affect their opinions. Researchers create new opinion models to develop insights into human interactions and their effects on social phenomena. These models are far simpler than how opinions change in reality. A major challenge in modeling opinions is evaluating models by comparing them with opinions in real-world data [1]. Researchers are actively trying to develop good ways to do this. However, despite the challenges, opinion models still help researchers test and learn how social interactions affect our opinions. As researchers develop more opinion models, we will learn more about how opinions change and how these changes impact human behavior.

## Acknowledgements

We thank our young readers (Claire Ashlock, Elyse Ashlock, Jaidan Bradley, Nia Chiou, Taryn Chiou, Zoe Chiou, and Viaan Rao) and their teachers and relatives (Lyndie Chiou, Christina Chow, Danielle Lyles, and Lori Ziegelmeier) for helpful comments. We also thank Antonio Scala and our young reviewers and editors for their helpful feedback. We sometimes used Chat GPT for suggestions of ways to simplify the text. GJL, JL, and MAP acknowledge support from the National Science Foundation (Grant Number 1922952) through the Algorithms for Threat Detection (ATD) program. GJL and JL also acknowledge support from the National Science Foundation (through Grant Number 1829071).

## Glossary

*Consensus:* When everyone has the same opinion, we say that they have reached a consensus.

*Fragmented opinions:* A group of people have fragmented opinions when they are not in consensus. There are two or more different opinions.

*Mathematical model:* A description of something using mathematical rules and language. An example is a voter model of opinions and how they change.

*Social Network:* A collection of people and the social relationships between them. Examples of social relationships include friendships, family relationships, and social-media connections.

*Node:* An object in a social network that represents a person or other individual. Two nodes that have an edge between them are called “neighbors”.

*Edge:* A connection between two nodes in a social network. An edge can represent a friendship or some other social relationship.

*Update rule:* A rule that determines how people change their opinions in an opinion model. An update rule determines who interacts with each other and how people change their opinions when they interact.

*Voter model:* An opinion model in which pairs of people (or other individuals) are chosen randomly to agree on one of their opinions. There are many variations of voter models.

**Authors**

**Grace J. Li:** Grace is a postdoc at Lawrence Livermore National Laboratory. She earned her Ph.D. from the Department of Mathematics at UCLA in 2024. Her research interests are network science and data science. In addition to building and coding mathematical models, Grace likes arts and crafts and building physical models and miniatures. She also likes cooking, as well as convincing others to try her foods and share their opinions.

**Jiajie (Jerry) Luo:** Jerry earned his Ph.D. from the Department of Mathematics at UCLA in 2024. His research interests are topological data analysis and social systems. When Jerry is not doing mathematics, he likes to be outdoors, especially in the mountains.

**Kaiyan Peng:** Kaiyan works at Meta. She earned her Ph.D. from the Department of Mathematics at UCLA in 2022. Her Ph.D. research included work on opinion models, disease models that are coupled to opinion models, and a model of illegal logging.

**Mason A. Porter:** Mason is a professor in the Department of Mathematics at UCLA. He also has a 0% appointment in the Department of Sociology at UCLA and is an external faculty member of the Santa Fe Institute. In addition to studying networks and other topics in mathematics and its applications, Mason is a big fan of games of all kinds, fantasy, baseball (Go Dodgers!), the 1980s, and other delightful things. Mason has strong opinions on many things, including models of opinion dynamics. In opinion models, Mason is typically one of the stubborn nodes that seldom changes its opinion.